\documentclass{aa}
\usepackage{graphics}

\begin{document}      

   \title{NGC~4654: gravitational interaction or ram pressure stripping?}

   \author{B.~Vollmer}

   \offprints{B.~Vollmer, e-mail: bvollmer@mpifr-bonn.mpg.de}

   \institute{Max-Planck-Institut f\"ur Radioastronomie, Auf dem H\"ugel 69,
              D-53121 Bonn, Germany. 
              }

   \date{Received / Accepted}

   \authorrunning{B.~Vollmer}
   \titlerunning{NGC~4654}

\abstract{
The Virgo cluster spiral galaxy NGC~4654 is supposed to be a
good candidate for ongoing ram pressure stripping based on its
very asymmetric H{\sc i} distribution. However, this galaxy also
shows an asymmetric stellar distribution. Numerical simulations using 
ram pressure as the only perturbation can produce a tail structure of 
the gas content, but cannot account for its kinematical structure.
It is shown that a strong edge--on stripping event can produce
an asymmetric stellar distribution up to 800~Myr after the
stripping event, i.e. the galaxy's closest passage to the cluster 
center. Simulations using a gravitational interaction with the companion
galaxy NGC~4639 can account for the asymmetric stellar distribution
of NGC~4654, but cannot reproduce the observed extended gas tail.
Only a mixed interaction, gravitational and ram pressure, can 
reproduce all observed properties of NGC~4654. It is concluded that
NGC~4654 had a tidal interaction $\sim$500~Myr ago and is continuing to 
experience ram pressure. 
\keywords{
Galaxies: individual: NGC~4654 -- Galaxies: interactions -- Galaxies: ISM
-- Galaxies: kinematics and dynamics
}
}

\maketitle

\section{Introduction}

Maps of the gas content of spiral galaxies in the Virgo cluster have revealed 
that the H{\sc i} disks of cluster spirals are disturbed and considerably 
reduced (Cayatte et al. 1990, 1994). Despite their H{\sc i} deficiency,
these spiral galaxies do not show a lack of molecular gas (Kenney \& Young 1989).
These observational results suggest that 
gas removal due to the rapid motion of the galaxy within the intracluster
medium (ICM) (ram--pressure stripping; Gunn \& Gott 1972) is responsible for 
the H{\sc i} deficiency and the disturbed gas disks of theses cluster spirals. 
Nevertheless, it is still an open question where and when these galaxies lost 
their atomic gas.

In order to study in detail how ram pressure acts, we are modeling 
galaxy orbits and use a three-dimensional N--body code to simulate the 
gas kinematics of a spiral galaxy falling into the Virgo cluster 
(Vollmer et al. 2001). Ram pressure exerted by the ICM on the ISM of a
rapidly moving galaxy is explicitly included in this code.

One candidate for an ISM--ICM interaction in the Virgo cluster is NGC~4654. 
Its physical parameters are listed in Tab.~\ref{tab:parameters}.
Phookun \& Mundy (1995) observed this galaxy in the 21--cm--line with a high 
sensitivity ($\sim 10^{19}$~cm$^{-2}$) at the VLA. They found a very asymmetric 
H{\sc i} distribution with a compressed edge on one side and a long tenuous tail 
on the other side. They argued that this asymmetry is due to ram pressure
stripping that pushes the ISM beyond the galaxy's optical radius.

\begin{table}
      \caption{Physical Parameters of NGC~4654}
         \label{tab:parameters}
      \[
         \begin{array}{lr}
            \hline
            \noalign{\smallskip}
        {\rm Other\ names} &  {\rm UGC~7902} \\
                & {\rm VCC~1987} \\
                & {\rm CGCG~071-019}  \\
        $$\alpha$$\ (2000)$$^{\rm a}$$ &  12$$^{\rm h}43^{\rm m}56.6^{\rm s}$$\\
        $$\delta$$\ (2000)$$^{\rm a}$$ &  13$$^{\rm o}07'36''$$\\
        {\rm Morphological\ type}$$^{\rm a}$$ & {\rm SBcd} \\
        {\rm Distance\ to\ the\ cluster\ center}\ ($$^{\rm o}$$) & 3.4\\
        {\rm Optical\ diameter\ D}_{25}$$^{\rm a}$$\ ($$'$$) & 4.9\\
        {\rm B}$$_{T}^{0}$$$$^{\rm a}$$ & 10.75\\ 
        {\rm Systemic\ heliocentric\ velocity}$$^{\rm a}$$\ {\rm (km\,s}$$^{-1}$$)\ & 1035$$\pm$$3\\
        {\rm Distance\ D\ (Mpc)} & 17 \\
        {\rm Vrot}$$_{\rm max}\ {\rm (km\,s}$$^{-1}$$) & 130$$^{\rm b}$$,\ 139$$^{\rm c}$$ \\
        {\rm PA} & 121$$^{\rm o}$$\ $$^{\rm b}$$,\ 128$$^{\rm o}$$\ $$^{\rm c}$$\\
        {\rm Inclination\ angle} & 49$$^{\rm o}$$\ $$^{\rm b}$$,\ 51$$^{\rm o}$$\ $$^{\rm c}$$ \\
        {\rm HI\ deficiency}^{\rm d}$$ &  0.17$$\pm$$0.2\\
        \noalign{\smallskip}
        \hline
        \end{array}
      \]
\begin{list}{}{}
\item[$^{\rm{a}}$] RC3, de Vaucouleurs et al. (1991)
\item[$^{\rm{b}}$] Guharthakurta et al. (1988)
\item[$^{\rm{c}}$] Sperandio et al. (1995)
\item[$^{\rm{d}}$] Cayatte et al. (1994)
\end{list}
\end{table}

However, NGC~4654 does not only show an asymmetric gas disk, but also
has an asymmetric stellar distribution. This might be due to a
gravitational interaction with the nearby spiral galaxy NGC~4639.

In order to investigate the possibilities of tidal or ISM--ISM interaction, 
the code presented in Vollmer et al. (2001) has been extended and contains
now a non--collisional component. This gives the opportunity to
make simulations of a gravitational interaction with or without
additional ram pressure.

In this article I compare snapshots of four different simulations with
H{\sc i} observations of NGC~4654:
\begin{enumerate}
\item
ram pressure, analytically given gravitational potential of the galaxy;
\item
ram pressure, galactic halo, bulge, and disk simulated by non--collisional
particles;
\item
gravitational interaction;
\item
gravitational interaction and ram pressure.
\end{enumerate} 
The comparison of the gas distribution {\it and} velocity field
permits to discriminate between the different scenarios.

The plan of this article is the following: the numerical code is 
described in Sect.~2. The simulations using ram pressure as the
only perturbation and their comparison with observations are presented 
in Sect.~4. In Sect.~5 a scenario
of active stripping is discussed. The simulations using a gravitational
interaction with and without ram pressure and their comparison
with observations are presented in Sect.~6, followed by a discussion
of the results of all simulations in Sect.~7.
A summary and conclusions are given in Sect.~8.

\section{The model}

\subsection{The collisional component}

Since the model is described in detail in Vollmer et al. (2001), I will
only briefly summarize its main features.
The particles represent gas cloud complexes which are 
evolving in an analytically given gravitational potential of the galaxy.

10\,000 particles of different masses are rotating within this gravitational 
potential. The total gas mass is $M_{\rm gas}^{\rm tot}=5.8\,10^{9}$~M$_{\odot}$,
which is $\sim$30\% larger than the observed gas mass (Huchtmeier \& Richter 1989).
To each particle a radius is attributed depending on its mass. 
During the disk evolution the particles can have inelastic collisions, 
the outcome of which (coalescence, mass exchange, or fragmentation) 
is simplified following Wiegel (1994). 
This results in an effective gas viscosity in the disk. 

As the galaxy moves through the ICM, its clouds are accelerated by
ram pressure. Within the galaxy's inertial system its clouds
are exposed to a wind coming from the opposite direction of the galaxy's 
motion through the ICM. 
The temporal ram pressure profile has the form of a Lorentzian,
which is realistic for galaxies on highly eccentric orbits within the
Virgo cluster (Vollmer et al. 2001).
The effect of ram pressure on the clouds is simulated by an additional
force on the clouds in the wind direction. Only clouds which
are not protected against the wind by other clouds are affected.

\subsection{The analytical gravitational potential}

The fixed, analytical gravitational potential consists of three parts: 
the dark matter halo,
the stellar bulge, and the disk potential (Allen \& Santill\'an 1991). 
Using their definitions, the model parameters are: (i) halo: 
$a_{3}$=12~kpc (effective radius), $M_{3}=8.6\,10^{10}$~M$_{\odot}$
(halo mass), (ii) bulge: $b_{1}$=390~pc (bulge core radius), 
$M_{1}=5.6\,10^{9}$~M$_{\odot}$ (bulge mass), (iii) disk: $a_{2}$=2.7~kpc 
(disk scale length), $b_{2}$=250~pc (disk scale height), 
$M_{2}=2.6\,10^{10}$~M$_{\odot}$ (mass of the stellar disk).
The resulting velocity field has a constant rotation 
curve of $v_{\rm rot} \sim$150 km\,s$^{-1}$, which is
consistent with the rotation curves derived by Guharthakurta et al. (1988)
and Sperandio et al. (1995).

\subsection{The non--collisional component}

The non--collisional component consists of 49\,125 particles, which simulate
the galactic halo, bulge, and disk.
The characteristics of the different galactic components are shown in
Tab.~\ref{tab:param}.
\begin{table}
      \caption{Number of particles $N$, particle mass $M$, and smoothing
        length $l$ for the different galactic components.}
         \label{tab:param}
      \[
         \begin{array}{llll}
           \hline
           \noalign{\smallskip}
           {\rm component} & N & M\ ({\rm M}$$_{\odot}$$) & l\ ({\rm pc}) \\
           \hline
           {\rm halo} & 16384 & $$9.2\,10^{6}$$ & 1200 \\
           {\rm bulge} & 16384 & $$3.2\,10^{5}$$ & 180 \\
           {\rm disk} & 16384 & $$1.6\,10^{6}$$ & 240 \\
	   {\rm companion} & 11000 & $$9.2\,10^{6}$$ & 1200 \\
           \noalign{\smallskip}
        \hline
        \end{array}
      \]
\end{table}
The particle trajectories are integrated using an adaptive timestep for
each particle. This method is described in Springel et al. (2001).
The following criterion for an individual timestep is applied:
\begin{equation}
\Delta t_{\rm i} = \frac{20~{\rm km\,s}^{-1}}{a_{\rm i}}\ ,
\end{equation}
where $a_{i}$ is the acceleration of the particle i.
The minimum of all $t_{\rm i}$ defines the global timestep used 
for the Burlisch--Stoer integrator that integrates the collisional
component.

The setup of the initial conditions was made by the program described in
Hernquist (1993) with the following parameters:
\begin{itemize}
\item
halo: mass $M=1.5\,10^{11}$~M$_{\odot}$, core radius $r_{\rm c}$=3~kpc, tidal radius
$r_{\rm t}$=30~kpc, cutoff radius $R$=90~kpc,
\item
bulge: mass $M=5.2\,10^{9}$~M$_{\odot}$, scale length $l$=0.3~kpc, cutoff radius
$R$=30~kpc,
\item
disk: mass $M=2.6\,10^{10}$~M$_{\odot}$, scale length $l$=3~kpc, cutoff radius
$R$=45~kpc, disk thickness $z_{0}$=600~pc.
\end{itemize}

10\,000 collisional particles with a 1/R column density profile
were added and the system was evolved during 2~Gyr in order to obtain 
a relaxed system. At the end of the simulation the difference between the 
total energy and the total angular momentum of the system was smaller 
than 0.5\% of their initial value. During the last Gyr of this simulation
the disk properties (disk height, surface density profile, density profile)
did not change within 5\%. 
The final state of this simulation was used as the initial state for the
here presented simulations.

The companion galaxy with a mass of $10^{11}$~M$_{\odot}$ is simulated by 11\,000 
particles forming a Plummer sphere with with a core radius of 2~kpc.

\section{The simulations}

I have made four different simulations using different interactions and
different methods to model the galactic components (dark halo, stars, and
gas).
\begin{enumerate}
\item
Time dependent ram pressure stripping, no gravitational interaction.
The gravitational potential of the galaxy is fixed and analytically given.
\item
Time dependent ram pressure stripping, no gravitational interaction.
The gravitational potential of the galaxy is simulated by non--collisional 
particles.
\item
Gravitational interaction, no ram pressure stripping.
The gravitational potential of the galaxy is simulated by non--collisional 
particles.
\item
Gravitational interaction and constant ram pressure.
The gravitational potential of the galaxy is simulated by non--collisional 
particles.
\end{enumerate}

\section{Time dependent ram pressure stripping}

In Vollmer et al. (2001) we simulated different galaxy orbits
with different inclination angles $i$ between the orbital and the disk plane.
Phookun \& Mundy (1995) suggested that NGC~4654 is moving edge--on through
the ICM. Indeed, only simulations with $i<20^{\rm o}$ show an extended 
tail several 10$^{8}$~yr after the galaxy's closest passage to the cluster center.
Furthermore, the simulation has to reproduce the observed H{\sc i} deficiency of 
NGC~4654. With $i=0^{\rm o}$ (edge--on stripping), a maximum ram pressure
$p_{\rm ram}^{\rm max}>5000$~cm$^{-3}$(km\,s$^{-1}$)$^{2}$ leads to
a final H{\sc i} deficiency $DEF \geq 0.05$. It turned out that the
best correspondence between the observed and the model H{\sc i} surface
density exists for $p_{\rm ram}=5000$~cm$^{-3}$(km\,s$^{-1}$)$^{2}$.
In Vollmer et al. (2001) we have shown that the highest realistic
maximum ram pressure is 
$p_{\rm ram}^{\rm max} \sim 10000$~cm$^{-3}$(km\,s$^{-1}$)$^{2}$.
In order to obtain higher values the galaxy would have to approach M87
to distances where a damaging gravitational interaction is ineluctable.
Thus, values of the maximum ram pressure in the range
5000~cm$^{-3}$(km\,s$^{-1}$)$^{2} < p_{\rm ram} < 10000$~cm$^{-3}$(km\,s$^{-1}$)$^{2}$
and the inclination angles in the range of $0^{\rm o} \leq i \leq 10^{\rm o}$
can reproduce the H{\sc i} observations.
Since the observed H{\sc i} deficiency has an error of $\pm$0.2, the
final model deficiency of $DEF$=0.05 is consistent with the observed 
value ($DEF$=0.17), i.e. no H{\sc i} deficiency.
The best fit model parameters are given in Tab.~\ref{tab:modelparameters}.
\begin{table}
      \caption{Model parameters for time dependent ram pressure stripping}
         \label{tab:modelparameters}
      \[
         \begin{array}{lr}
            \hline
            \noalign{\smallskip}
        {\rm maximum\ ram pressure}\ \big({\rm cm}$$^{-3}$$({\rm km\,s}$$^{-1}$$)$$^{2}$$\big)\ \ &  5000\\
        {\rm inclination\ angle\ between\ orbital\ and\ disk\ plane} & 0$$^{\rm o}$$ \\
        {\rm final\ HI\ deficiency} &  0.1\\
        \noalign{\smallskip}
        \hline
        \end{array}
      \]
\end{table}

\subsection{The projection on the sky \label{sec:sky}}

In order to project the model gas distribution on the sky,
three angles are needed: the position angle, the inclination
angle, and the azimuthal angle within the plane of the galactic disk.
The position angle can be determined observationally.
If one assumes that the spiral arms of NGC~4654 are trailing,
its velocity field (Phookun \& Mundy 1995) places its northern
edge in front of the galaxy center, i.e. the sign of the
inclination angle $i$ is known. 

In general, galaxies have
trailing spiral arms. Galaxies with leading spiral arms
are assumed to have undergone a plunging retrograde encounter with 
another galaxy (see, e.g., Thomasson et al. 1989 or
Byrd et al. 1993), which should not be the case for NGC~4654.
Since NGC~4654 harbors a bar and two normal spiral arms, there is
no apparent reason why its spiral arms should be leading.
It will be shown in Sect.~\ref{sec:active} that the assumption
of trailing spirals is important for the model.

With a given position angle (PA) and a given inclination angle $i$,
only the azimuthal ($\alpha$) in the galactic plane can be varied. 
I assume that the galaxy orbit within the cluster is 
approximately linear in space (see Vollmer et al. 2001). If NGC~4654 
is emerging from the cluster core, it must be located in front of M87,
because its radial velocity with respect to the cluster mean
is negative ($\Delta v \sim -100$~km\,s$^{-1}$). Furthermore,
the component of its three--dimensional
velocity vector parallel to the right ascension axis
must be negative (the galaxy is moving to the east).

\begin{figure}
        \resizebox{\hsize}{!}{\includegraphics{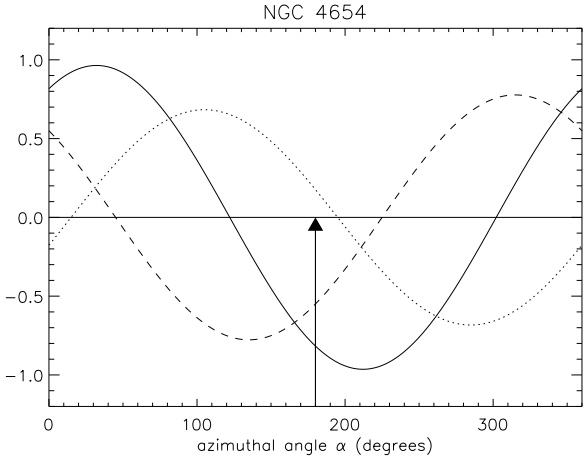}}
        \caption{The three components of the galaxy's
          velocity vector ${\bf v^{\rm gal}}$ as a function of the 
          azimuthal angle $\alpha$.
          Solid: $v^{\rm gal}_{1}$, dotted: $v^{\rm gal}_{2}$, 
          dashed: $v^{\rm gal}_{3}$. The chosen angle is marked by an arrow. 
          $v^{\rm gal}_{1}$ is the component parallel 
          to the axis of right ascension (positive is to negative right 
          ascensions), $v^{\rm gal}_{2}$ is that parallel to the axis of 
          declination, and $v^{\rm gal}_{3}$ is that parallel to the line 
          of sight.
        } \label{fig:winkel}
\end{figure}

In Fig.~\ref{fig:winkel} the three components of the galaxy's
velocity vector ${\bf v^{\rm gal}}$ are shown as a function of $\alpha$.
$v^{\rm gal}_{1}$ is the component parallel to the axis of right
ascension (positive is to negative right ascensions), $v^{\rm gal}_{2}$ 
is that parallel to the axis of declination, and $v^{\rm gal}_{3}$ is
that parallel to the line of sight.
The constraints $v^{\rm gal}_{3} < 0$~km\,s$^{-1}$ and 
$v^{\rm gal}_{1} < 0$~km\,s$^{-1}$ imply 
$130^{\rm o} \leq \alpha \leq 220^{\rm o}$.
It turned out that $\alpha=180^{\rm o}$ leads to the best fit to observations.
This angle is marked by an arrow in Fig.~\ref{fig:winkel}.

\subsection{The simulations \label{sec:simulation}}

\begin{figure}
        \resizebox{8cm}{!}{\includegraphics{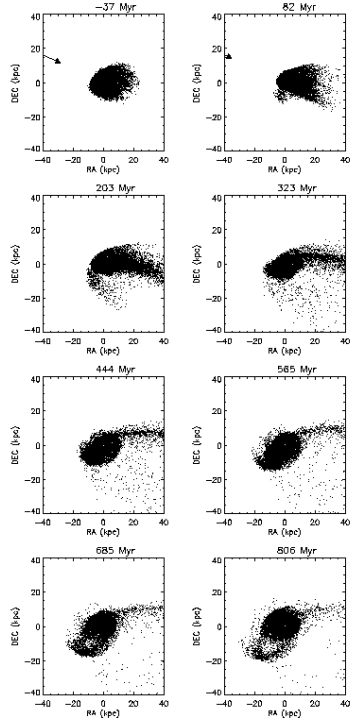}}
        \caption{Snapshots of the evolution of the galaxy's ISM with a fixed
        gravitational potential.
        The $x$-axis corresponds to the right ascension,
        The $y$-axis to declination. 
        The elapsed time is indicated at the top of each panel.
        The position and inclination angle of NGC~4654 are used.
        The galaxy rotates counter--clockwise. It is moving to the 
        north-east, i.e. the wind is coming from the north-east
        (indicated by the arrows). The length of the arrow is proportional 
        to the ram pressure ($\rho_{\rm ICM} v_{\rm gal}^{2}$).
        } \label{fig:n4654_simulation}
\end{figure}

\begin{figure}
        \resizebox{8cm}{!}{\includegraphics{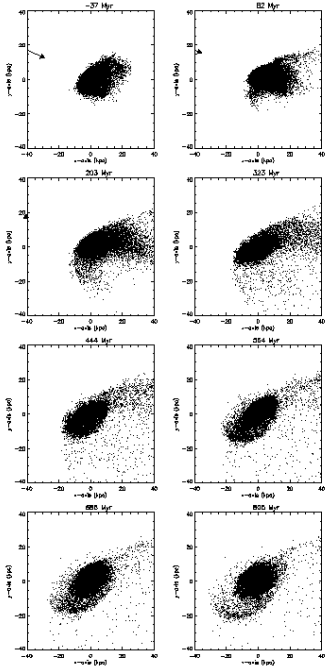}}
        \caption{Snapshots of the evolution of the galaxy's ISM for a galaxy model
        including a non--collisional component.
        The $x$-axis corresponds to the right ascension,
        The $y$-axis to declination. 
        The elapsed time is indicated at the top of each panel.
        The position and inclination angle of NGC~4654 are used.
        The galaxy rotates counter--clockwise. It is moving to the 
        north-east, i.e. the wind is coming from the north-east
        (indicated by the arrows). The length of the arrow is proportional 
        to the ram pressure ($\rho_{\rm ICM} v_{\rm gal}^{2}$).
        } \label{fig:n4654_simulation_new}
\end{figure}

Fig.~\ref{fig:n4654_simulation} and \ref{fig:n4654_simulation_new}
show the evolution of the galaxy's ISM after its closest passage to the cluster center
for the model with a fixed gravitational potential and
a galaxy model including a non--collisional component respectively.
The position and inclination angle of NGC~4654 are used.
The time $t$ can be found above each frame. The closest passage
to the cluster center corresponds to $t$=0~yr. 
The arrow indicates the direction
of the wind, i.e. it is opposite to the galaxy's motion within the cluster.
The length of the arrows is proportional to the ram pressure
$p_{\rm ram}=\rho_{\rm ICM}v_{\rm gal}^{2}$, where $\rho_{\rm ICM}$
is the ICM density and $v_{\rm gal}$ is the galaxy velocity with 
respect to the cluster mean. The northern edge of the galaxy is in front 
of the galaxy center. The galaxy is rotating counter--clockwise.
Since both simulations are very similar, I will describe in the following
the main features for both simulations:
at approximately maximum ram pressure ($t=0$) an overdensity builds up
in the direction of the galaxy's motion, where the gas has been
pushed to smaller radii. At $t\sim 100$~Myr the wind has driven out
the gas to the north where the vector of the galactic rotation
velocity is parallel to the wind direction. In Vollmer et al. (2001)
this is called the accelerated arm. A weak decelerated arm can
also be seen in the south. At $t\sim 200$~Myr ram pressure has already
ceased completely and the evolution of the galaxy is entirely
due to rotation and re--accretion of the gas, which has not been 
accelerated to the escape velocity ($v_{\rm esc} \sim \sqrt{2} v_{\rm rot}$).
Due to rotation, the accelerated arm moves to the north
and a part of its material falls back onto the galaxy.
At $t > 350$~Myr an asymmetric shell structure is building up in the
south, which is most prominent in the south--east. This south--eastern part 
expands and forms an extended tail at $t \sim 700$~Myr after the
galaxy has passed the cluster center.

The main difference of the gas dynamics
between the two models is the that the gravitational
potential of model galaxy including a non--collisional component can change
with time. The ISM--ICM interaction pushes the gas to smaller
galactic radii and heats it. This process also heats the stellar disk.
At maximum ram pressure the surface density in the north--east
increase by a factor $\sim$2 and $\sim$2.5\,10$^{9}$~M$_{\odot}$ of gas, which
represents $\sim$10\% of the total enclosed mass within 5~kpc, are
displaced to the south--west. 
In addition, this gravitational perturbation triggers the formation of an
asymmetry of the stellar distribution, which can also heat the stellar disk.
Thus, the length scale of its gravitational potential in $z$
direction increases, i.e. the motion of the gas is less two
dimensional than in the case of the fixed gravitational potential.
The result is that the asymmetries of the gas distribution are smoothed out in the 
case of the model including a non--collisional component. 
In a real galaxy the gas might cool through shocks during the interaction.
Moreover, the disk heating through the the heated gas might be a numerical
artifact due to the relatively small number of particles used in the simulation.
It is not clear if the gravitational heating is sufficient to produce the large 
scale height observed in the model.

There is another big difference between the two models: the disk and
bulge stars react to the change of the gravitational potential 
when the gas is pushed to small galactic radii. 
This produces a gravitational shock, i.e.
the stellar surface density also shows an asymmetry with a delay
of $\sim$400~Myr. 
\begin{figure}
        \resizebox{\hsize}{!}{\includegraphics{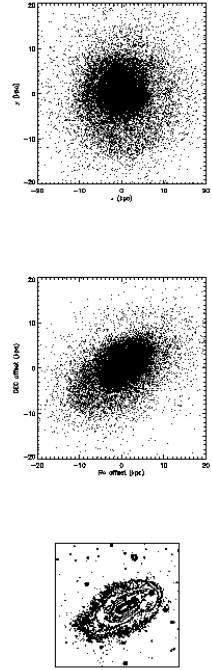}}
        \caption{Top panel: model distribution of the disk stars at $t$=800~Myr
          seen face--on.
          Middle panel: model distribution of the disk stars at $t$=800~Myr 
          projected on the sky with the PA and $i$ of NGC~4654.
          Bottom panel: K band image of NGC~4654 (Boselli et al. 1997).
        } \label{fig:stars_all}
\end{figure}
This asymmetric distribution rotates slowly and smears out within
a few rotation times. Fig.~\ref{fig:stars_all} shows the distribution
of disk stars seen face--on and projected on the sky with the
parameters of NGC~4654 at $t$=800~Myr, i.e. the snapshot that I want to
compare with observations. The bottom panel of Fig.~\ref{fig:stars_all}
shows the K band image of Boselli et al. (1997).
Seen face--on, the stellar surface density of the outer disk ($R > 6$~kpc)
is higher in the north than in the south. The stellar density in the
inner disk is high with a strong decline at $R \sim 6$~kpc.
This translates into an elliptical projected stellar distribution with
the galaxy center in the south--eastern focal point.
This resembles the overall observed distribution of the old stellar
component in K band (Fig.~\ref{fig:stars_all} bottom panel).
Thus the observed overall asymmetry of the stellar distribution can
be reproduced by a gravitational shock due to edge--on ram pressure
stripping, but it is not possible to create a bar nor spiral arms in 
this way. 

\begin{figure}
        \resizebox{\hsize}{!}{\includegraphics{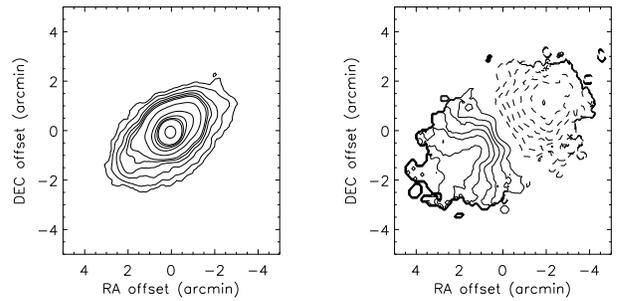}}
        \caption{Left panel: model stellar distribution at $t$=800~Myr.
          Right panel: model stellar velocity field.
          Both distributions are projected on the sky using PA and $i$
          of NGC~4654.
        } \label{fig:stars}
\end{figure}

Fig.~\ref{fig:stars} shows the stellar distribution and its velocity
field at $t$=800~Myr projected on the sky with the parameters of NGC~4654.
The velocity field shows clearly a steepening in the north--west,
whereas the southern part of the galaxy has the velocity field
of a rotation curve that flattens in the outer disk.

\subsection{Comparison between observations and simulations \label{sec:compare}}

In this Section the gas distribution and the velocity field
of NGC~4654 are directly compared to the last model snapshots of
Fig.~\ref{fig:n4654_simulation} and \ref{fig:n4654_simulation_new},
i.e. at $t$=800~Myr.

\begin{figure}
        \resizebox{6cm}{!}{\includegraphics{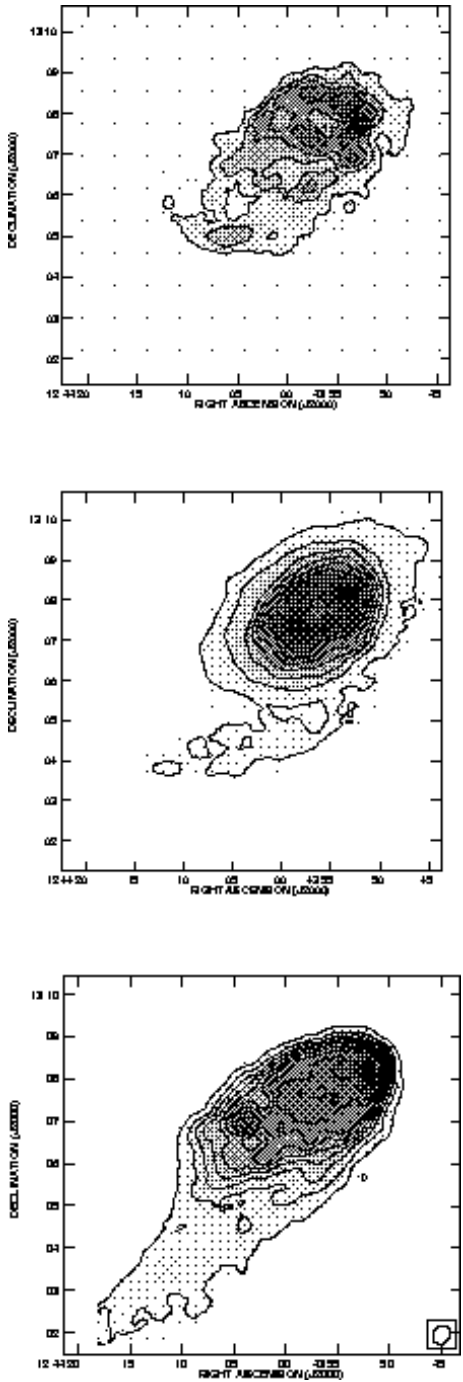}}
        \caption{Top panel: model distribution of the H{\sc i} gas
          using a fixed gravitational potential.
        The contours correspond to 
        (1, 5 ,9 ,13, 17, 21, 25)$\times$10$^{19}$~cm$^{-2}$.
        Middle panel: distribution of the H{\sc i} gas
        using the model including a non--collisional component
        with the same contour levels.
        Bottom panel: H{\sc i} distribution of NGC~4654 (Phookun \& Mundy 1995)
        with the same contour levels.
        } \label{fig:HI_distributions}
\end{figure}

Fig.~\ref{fig:HI_distributions} shows the H{\sc i} distribution of
NGC~4654 (bottom panel) together with the model gas distributions convolved
to the beamsize of the 21--cm--line observations (25$''$).
The top panel shows the gas distribution of the model
with a fixed gravitational potential, the middle panel shows
that of the model  including a non--collisional component.
The main characteristics of the observed gas distribution 
(Fig.~\ref{fig:HI_distributions} bottom panel) are
(i) the prominent maximum at the north--western edge of the disk,
(ii) the overall asymmetry within the disk with more gas
in the south--east, (iii) the linear extended tail to the south--east. 
The model using a fixed gravitational potential (Fig.~\ref{fig:HI_distributions} 
top panel) can reproduce feature (i) and (ii). 
It also shows an extended tail to the south--east, but it has a higher column
density, is curved, and is much less extended than the observed tail.
The model including a non--collisional component (Fig.~\ref{fig:HI_distributions} 
middle panel) fails in reproducing feature (i) and (ii), but 
can reproduce feature (iii), i.e. the extended, low column density, 
south--eastern tail. However, the model tail is slightly curved.
The low column density of this tail is due to the tidal heating of the
disk during the ICM--ISM interaction (see Sect.~\ref{sec:simulation}).

\begin{figure}
        \resizebox{6cm}{!}{\includegraphics{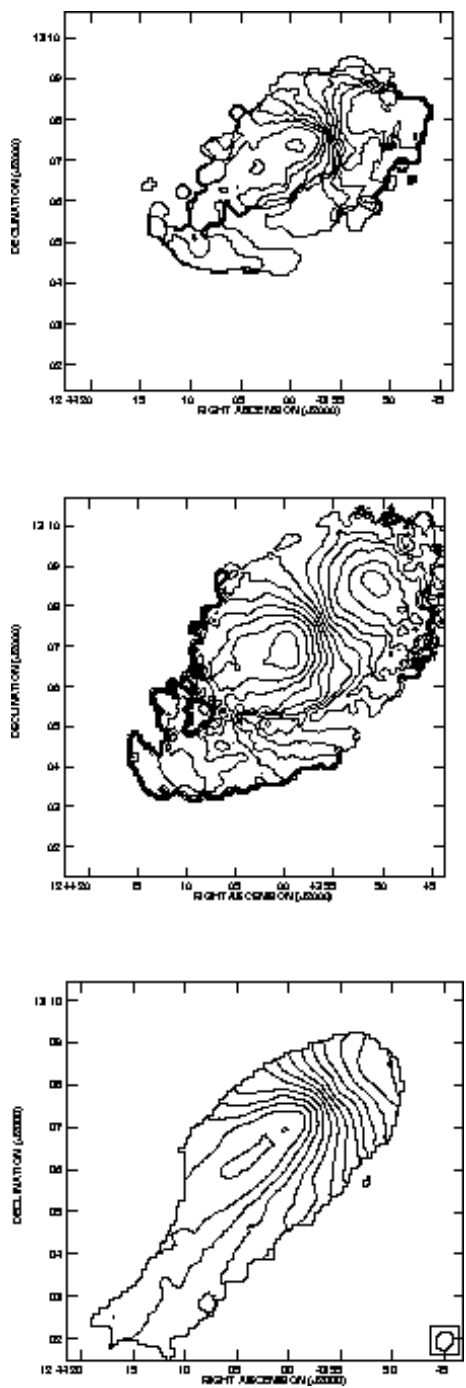}}
        \caption{Top panel: model H{\sc i} velocity field using a fixed gravitational
          potential. The contours, in km\,s$^{-1}$, are from 900 (north--west)
        to 1160 (south--east) in steps of 20 for both velocity fields.
        Middle panel: gas velocity field using the model including a 
        non--collisional component with the same contour levels.
        Bottom panel : H{\sc i} velocity field of NGC~4654 (Phookun \& Mundy 1995)
         with the same contour levels.
        } \label{fig:HI_velfields}
\end{figure}

The H{\sc i} velocity fields of the two models and the observed velocity
field can be seen in Fig.~\ref{fig:HI_velfields}. 
The main characteristics of the observed H{\sc i} velocity field
(Fig.~\ref{fig:HI_velfields} bottom panel) are (i) the plateau of constant rotation
velocity in the south--east, (ii) the steepening of the rotation curve
in the north west, (iii) the structure of the velocity in the extended
south--eastern tail.
The model using a fixed gravitational potential (Fig.~\ref{fig:HI_velfields} 
top panel) shows the features (i) and (ii), but fails in reproducing feature (iii).
The model including a non--collisional component (Fig.~\ref{fig:HI_velfields} 
middle panel) shows a less clear plateau, but feature (ii) is also
visible. It also fails in reproducing the observed velocity field in the
extended south--eastern tail.

\begin{figure}
        \resizebox{\hsize}{!}{\includegraphics{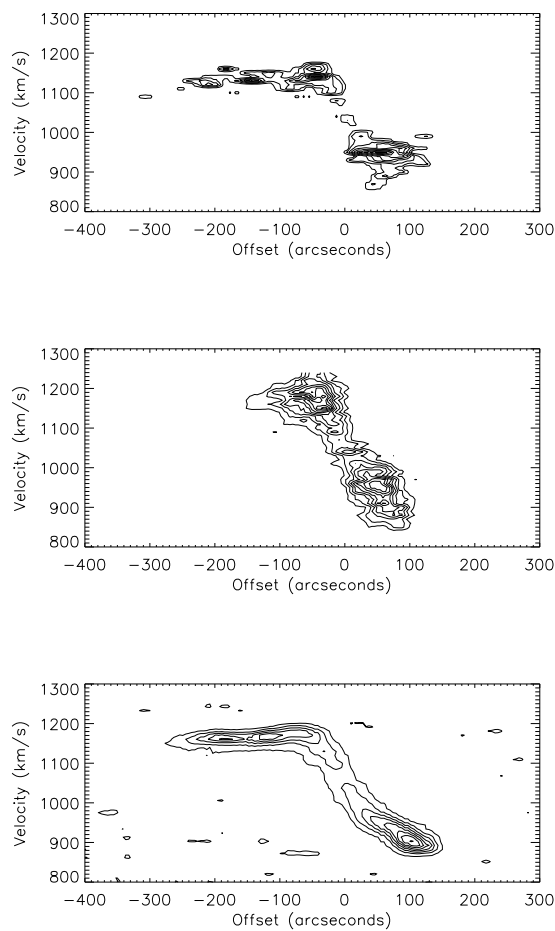}}
        \caption{Position--velocity cut through the H{\sc i} cube along the major axis.
        Top panel: model using a fixed gravitational potential.
        Middle panel: model including a non--collisional component.
        Bottom: H{\sc i} (Phookun \& Mundy 1995).
        } \label{fig:p-v_major}
\end{figure}

\begin{figure}
        \resizebox{\hsize}{!}{\includegraphics{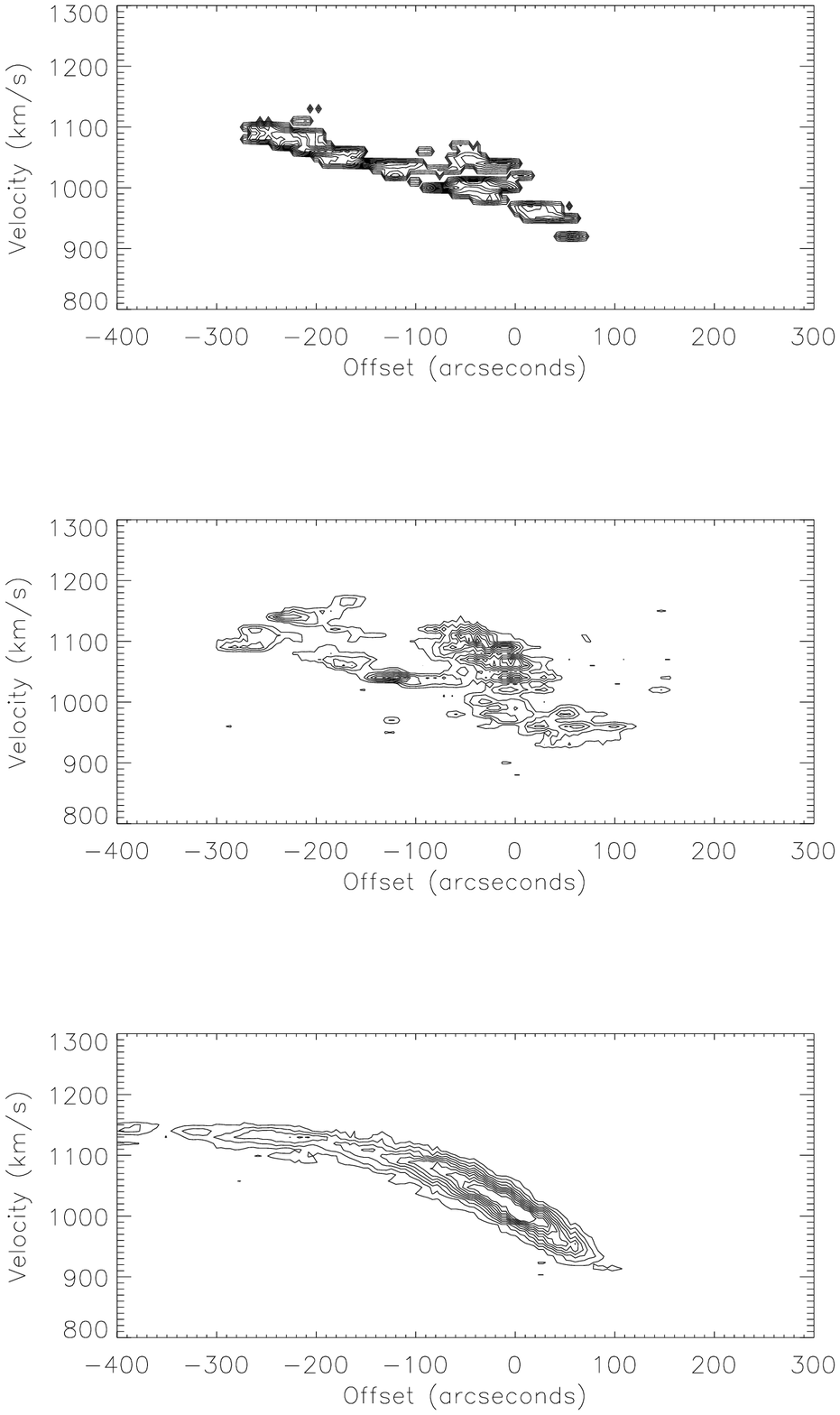}}
        \caption{Position--velocity cut through the H{\sc i} cube along the tail.
        Top panel: model using a fixed gravitational potential.
        Middle panel: model including a non--collisional component.
        Bottom: H{\sc i} (Phookun \& Mundy 1995).
        } \label{fig:p-v_tail}
\end{figure}

\begin{figure}
        \resizebox{\hsize}{!}{\includegraphics{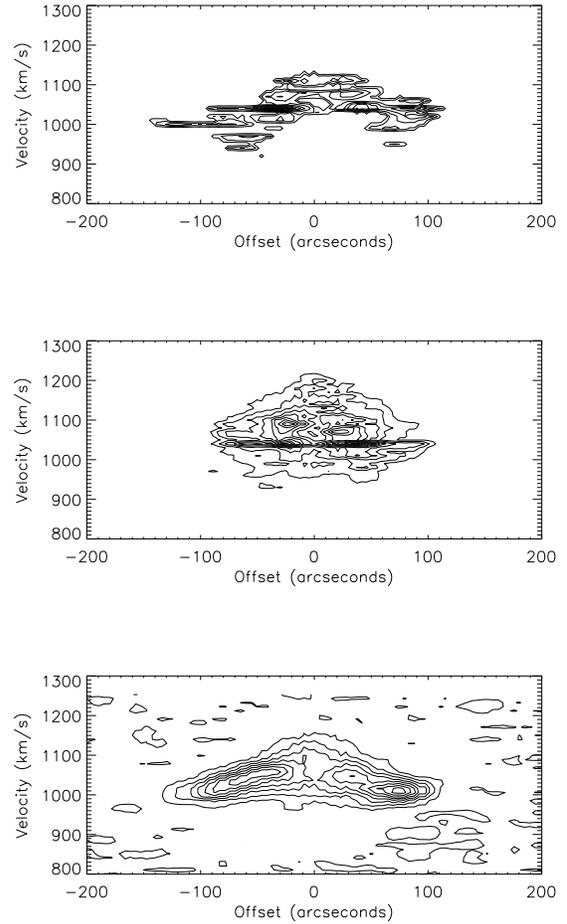}}
        \caption{Position--velocity cut through the H{\sc i} cube along the minor axis.
        Top panel: model using a fixed gravitational potential.
        Middle panel: model including a non--collisional component.
        Bottom: H{\sc i} (Phookun \& Mundy 1995). 
        } \label{fig:p-v_minor}
\end{figure}

In order to compare the characteristic structure of the H{\sc i} velocity fields, 
the position--velocity diagrams
along (i) the major axis, (ii) the tail, and (iii) the minor axis 
are shown in Fig.~\ref{fig:p-v_major}, \ref{fig:p-v_tail}, and
\ref{fig:p-v_minor}, respectively.
The model cut for a fixed gravitational potential along the major axis 
(Fig.~\ref{fig:p-v_major}, top panel) nicely shows the observed plateau
of constant rotation velocity. At the opposite side, the
main component is spatially reduced with respect to the side with positive 
offsets and shows a constant rotation.
Thus, this simulation is able to reproduce the south--eastern part of the
position--velocity diagram, but not the north-western part.

The cut for the model including a non--collisional component
(Fig.~\ref{fig:p-v_major}, middle panel) also shows a plateau at the
south--eastern side (negative offsets) that is spatially reduced
having only one third of the extent of the observed plateau.
The linear rise at the north--western side is reproduced.
At the north--western edge the rotation curve is falling again
producing closed contours in the velocity field. This is due to the
velocity dispersion of the gas within the disk, which is higher than
that of the observed galaxy, because there is no explicit cooling mechanism
included in the model.

The model cuts through the tail (Fig.~\ref{fig:p-v_tail},
top and middle panel) are very similar for both models.
As already seen above, these model cuts
differ from the observed one in the sense that the velocity
structure of the model tail is almost linear, whereas that of the
observed tail is significantly curved.

The model cuts along the minor axis (Fig.~\ref{fig:p-v_minor}) 
shows qualitatively the same deviations from circular rotation
as the cut through the observed data cube.

\section{The active stripping scenario \label{sec:active}}

The simulations of Sect.~\ref{sec:simulation} show that ram pressure
pushes gas beyond the optical radius of the galaxy at $t>0$~Myr.
A tail--like structure is built up at even later times, clearly
after the galaxy's closest passage to the cluster center.
This result rules out the scenario of active ram pressure stripping.

\begin{figure}
        \resizebox{\hsize}{!}{\includegraphics{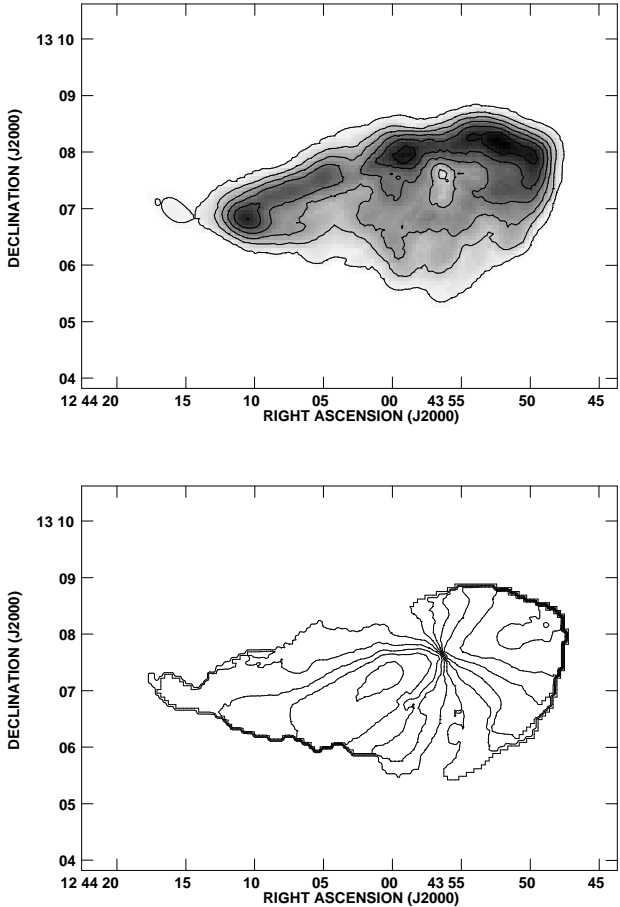}}
        \caption{Top: gas distribution of a simulation snapshot 
        at $t \sim 100$~Myr with
        a maximum ram pressure of 2000~cm$^{-3}$(km\,s$^{-1}$)$^{2}$
        and $i=0^{\rm o}$ (edge-on stripping). 
        The contours correspond to (1, 5 ,9 ,13, 17, 21, 25)$\times10^{19}$~cm$^{-2}$.
        Bottom: corresponding velocity field. The contours, in km\,s$^{-1}$, 
        are from 940 (north--west) to 1140 (south--east) in steps of 20.
        } \label{fig:HI_dist+vfield_61}
\end{figure}

Nevertheless, if one assumes that the efficiency of ram pressure 
to push the atomic gas has been greatly underestimated and that
the galaxy is moving through a region of enhanced ICM density,
it is possible to compare another simulation snapshot with the
observations, where ram pressure stripping is still active.
I have chosen the following
simulation parameters to best reproduce the observed
H{\sc i} distribution with the extended tail: maximum ram pressure
$p_{\rm ram}$=2000~cm$^{-3}$(km\,s$^{-1}$)$^{2}$, and $i=0^{\rm o}$ 
(edge--on stripping). The timestep of the snapshot is $t \sim 100$~Myr, 
i.e. ram pressure is still active.
The gas distribution and velocity field of this snapshot can be
seen in Fig.~\ref{fig:HI_dist+vfield_61}.
The tail represents the accelerated arm (cf. Fig.~\ref{fig:n4654_simulation}).
The assumption that the spiral arms are trailing places it
imperatively in the north (the galaxy rotates counter--clockwise).
The column density is uniform over the whole length of the tail, 
which is not observed. Concerning the velocity field, the southern plateau of
constant velocity is inclined with respect to the optical major axis.
Thus, this snapshot clearly does not fit the H{\sc i} observations
appropriately. 

If one goes still further and drops the assumption that the spiral 
arms of NGC~4654 are trailing (Sect.~\ref{sec:sky}), the accelerated arm
in Fig.~\ref{fig:HI_dist+vfield_61} would be located in the south
as observed. In this case, two discrepancies between
the model and the observations persist: (i) the uniform column density
of the model tail and (ii) the inclination of the southern plateau
of constant rotation velocity of the model.

\section{Gravitational interaction \label{sec:grav}}

\subsection{Gravitational interaction alone}

In this simulation we assume that NGC~4654 has been gravitationally
perturbed by its companion NGC~4639. Since we need a close encounter
in order to generate the observed asymmetry of the stellar disk of NGC~4654
without disrupting it, I choose a retrograde encounter.
The perturbing galaxy is modeled as a plummer sphere with a 
mass of $10^{11}$~M$_{\odot}$ and a length scale of 2~kpc.
The impact parameter is 21~kpc and the maximum relative velocity is
350~km\,s$^{-1}$. The minimum distance in the plane of the disk
of NGC~4654 is 18~kpc. I assume here that the inclination angle
of the disk of NGC~4639 is such that there is no ISM--ISM interaction.

I define the following coordinate system:
$x-y$ plane: disk plane, $z$ axis: perpendicular to the disk plane.
The origin is the center of NGC4654. In this system the initial conditions 
for NGC4639 are: $x$=-40~kpc, $y$=-70~kpc, $z$=70~kpc, $v_{x}$=0~km\,s$^{-1}$,
$v_{y}$=130~km\,s$^{-1}$, $v_{z}$=-130~km\,s$^{-1}$.
It is marked as a cross on the snapshots.

At $t$=1~Gyr, the projected distance and radial velocity difference
with respect to NGC~4654 using the PA and $i$ of
NGC~4654 are: $\tilde{x}$=60~kpc, $\tilde{y}$=55~kpc, 
$v_{\rm r}$=70~km\,s$^{-1}$. This compares to the observed projected 
position and radial velocity with respect to NGC~4654 of
$\tilde{x}_{\rm obs}$=80~kpc, $\tilde{y}_{\rm obs}$=39~kpc, 
$v_{\rm r}^{\rm obs}$=-35~km\,s$^{-1}$.
Both, positions and radial velocity, are close enough to suppose that the 
simulation should show the main characteristics of a possible gravitational 
interaction between NGC~4654 and NGC~4639.

\begin{figure*}
        \resizebox{\hsize}{!}{\includegraphics{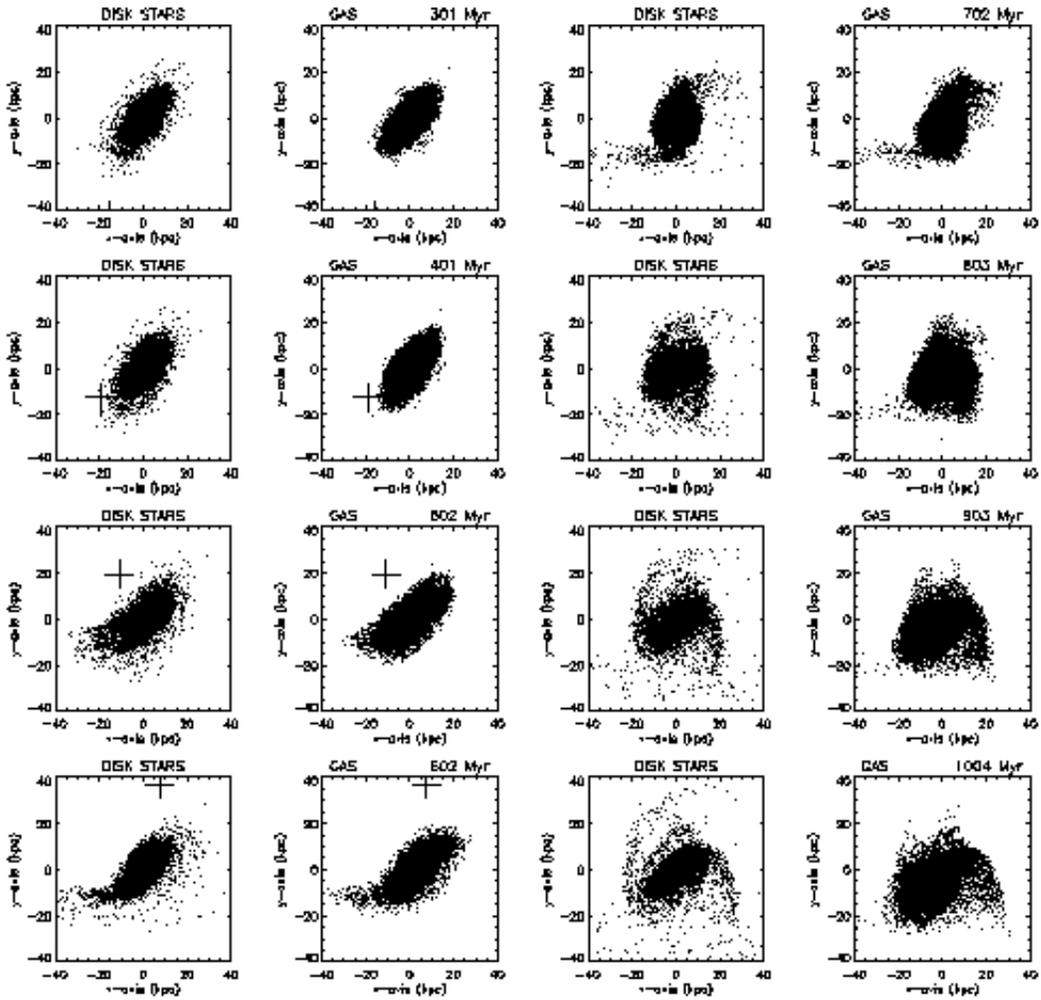}}
        \caption{Snapshots of the tidal interaction between NGC~4654
          and NGC~4639. The left panels show the stellar disk, the
          panels show the gas. The timesteps are marked on the
          upper right of each panel showing the gas.
          NGC~4639 is marked as a cross.
        } \label{fig:simulation_106}
\end{figure*}

\begin{figure}
        \resizebox{\hsize}{!}{\includegraphics{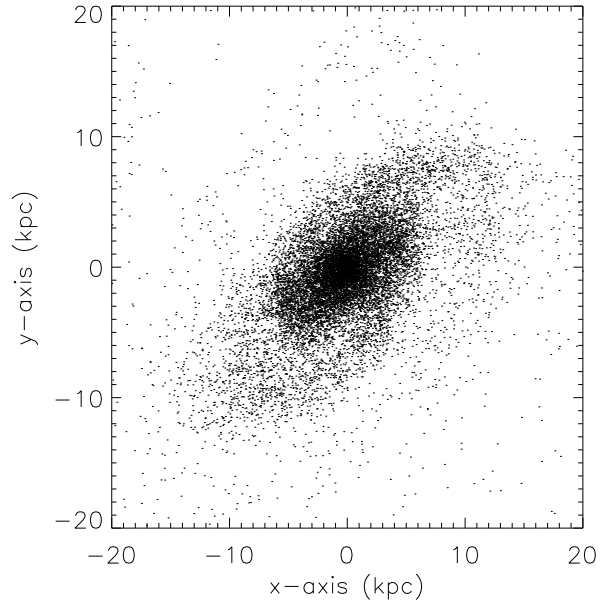}}
        \caption{Close--up of the stellar disk of the final snapshot
          after 1~Gyr (see Fig~\ref{fig:simulation_106}).
        } \label{fig:simulation_106_close}
\end{figure}

Fig.~\ref{fig:simulation_106} shows eight timesteps of this simulation.
The timestep of closest encounter is $t$=460~Myr. After the passage of
NGC~4639, NGC~4654 begins to form a large bar with two asymmetric 
spiral arm, the south--eastern being more extended than the north-western arm.
After $\sim$350~Myr these large scale spirals begin to wind up.
A small, negligible number of stellar particles is initially located outside the gas
disk. They are responsible for the low density outer spiral arms
for $t > 700$~Myr that are not present in the gas distribution.
On the other hand, the gas distribution has a $1/R$ profile, whereas
the stellar distribution has an exponential profile. This causes the
difference between the high column density structure of the gaseous
and stellar distribution.
In the inner part of the galaxy, i.e. within its optical radius,
a bar with two asymmetric spiral arms survive until $t$=1~Gyr, i.e.
$\sim$550~Myr after the closest encounter. The north--western spiral
arm is much tighter than the south-western one as observed for NGC~4654
in the optical.
Fig.~\ref{fig:simulation_106_close} shows a closer view of the stellar disk
at $t$=1~Gyr. 
The bar and the asymmetric spiral arms can be recognized.
This morphology resembles closely that observed in the K band
(Fig.~\ref{fig:stars_all}, bottom panel).

\subsection{Gravitational interaction and ram pressure}

As a next step, ram pressure is included in the model:  
$p_{\rm ram}=m_{\rm p}n_{\rm ICM}v_{\rm gal}^{2}$,
where $m_{\rm p}$ is the proton mass, $n_{\rm ICM}$ the density of the
intracluster medium, and $v_{\rm gal}$ the velocity of the galaxy
within the cluster. I have chosen it to be constant in order to keep 
the model as simple as possible. The values for the density
of the intracluster medium and the galaxy's velocity are typical
for a galaxy located at the projected distance of NGC~4654, i.e.
$n_{\rm ICM}=2\,10^{-4}$~cm$^{-2}$ and $v_{\rm gal}$=1000~km\,s$^{-1}$.
The direction of the galaxy's motion is parallel to its major axis
and is pointing to the north--west. The perturbing mass simulating
NGC~4639 has the same orbits as described in Sect.~\ref{sec:grav}.

\begin{figure*}
        \resizebox{\hsize}{!}{\includegraphics{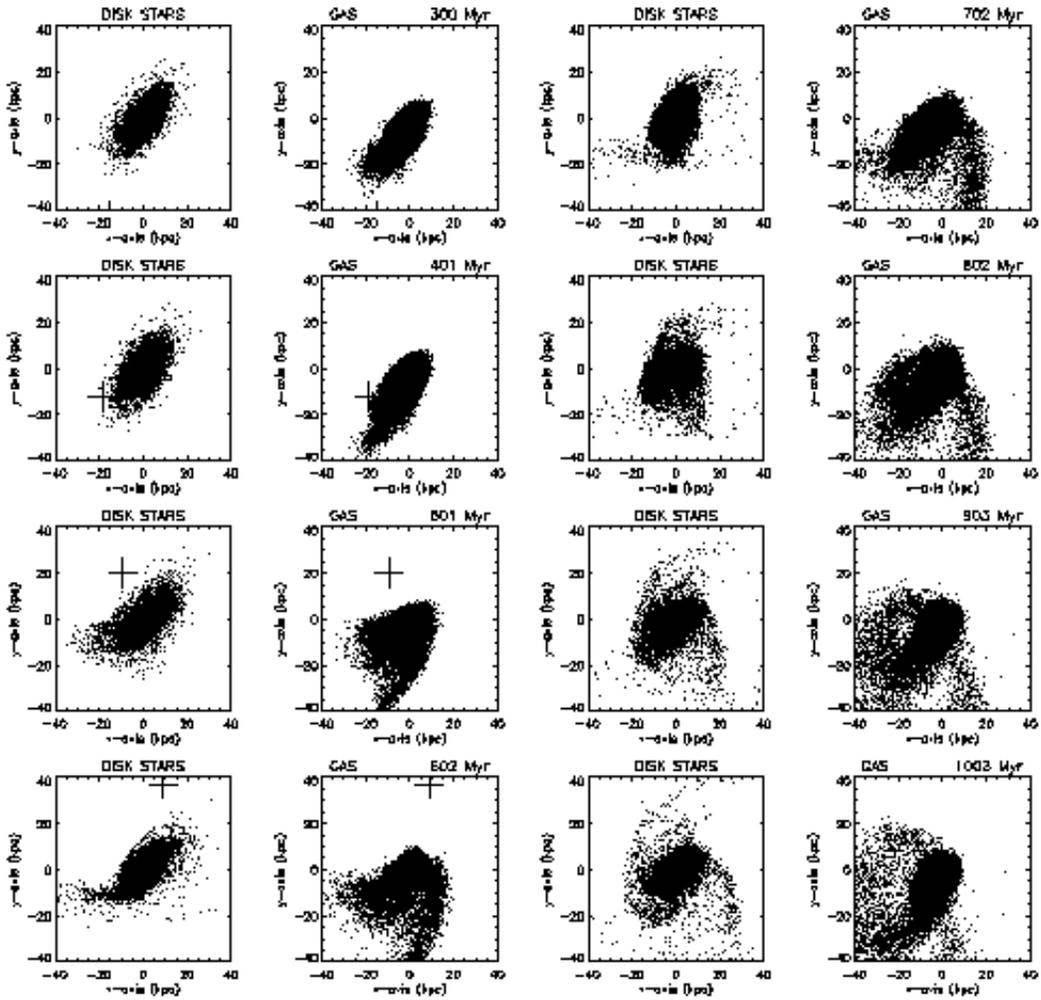}}
        \caption{Snapshots of the tidal interaction between NGC~4654
          and NGC~4639 including a constant ram pressure. 
          The left panels show the stellar disk, the
          panels show the gas. The timesteps are marked on the
          upper right of each panel showing the gas.
          NGC~4639 is marked as a cross.
        } \label{fig:simulation_110}
\end{figure*}

Fig.~\ref{fig:simulation_110} shows eight timesteps of this simulation.

Already at $t$=300~Myr ram pressure pushes the gas to the south--east
of the galaxy that has been forced to larger galactic radii due to the
tidal interaction with the perturbing mass. It forms a gaseous spiral arm
at the western side of the galaxy at $t$=400~Myr which then moves to the north.
It arrives at the northern edge of the galaxy at $t$=700~Myr and proceeds
further to the south--east, where it forms an extended tail structure at 
$t$=1~Gyr, which is corresponds to the observed tail of NGC~4654.
As expected, the stellar distributions at this final timestep for the
simulation with and without ram pressure (Fig.~\ref{fig:simulation_110}
and Fig.~\ref{fig:simulation_106}) are the same since the gas displacement
is not important enough to trigger a strong stellar asymmetry as observed
in Sect.~\ref{sec:compare}.

\subsection{Comparison between observations and simulations 
\label{sec:compare_grav}}

In this Section the gas distribution and the velocity field
of NGC~4654 are directly compared to the last model snapshots of
Fig.~\ref{fig:simulation_106} and \ref{fig:simulation_110},
i.e. at $t$=1~Gyr.

\begin{figure}
        \resizebox{6cm}{!}{\includegraphics{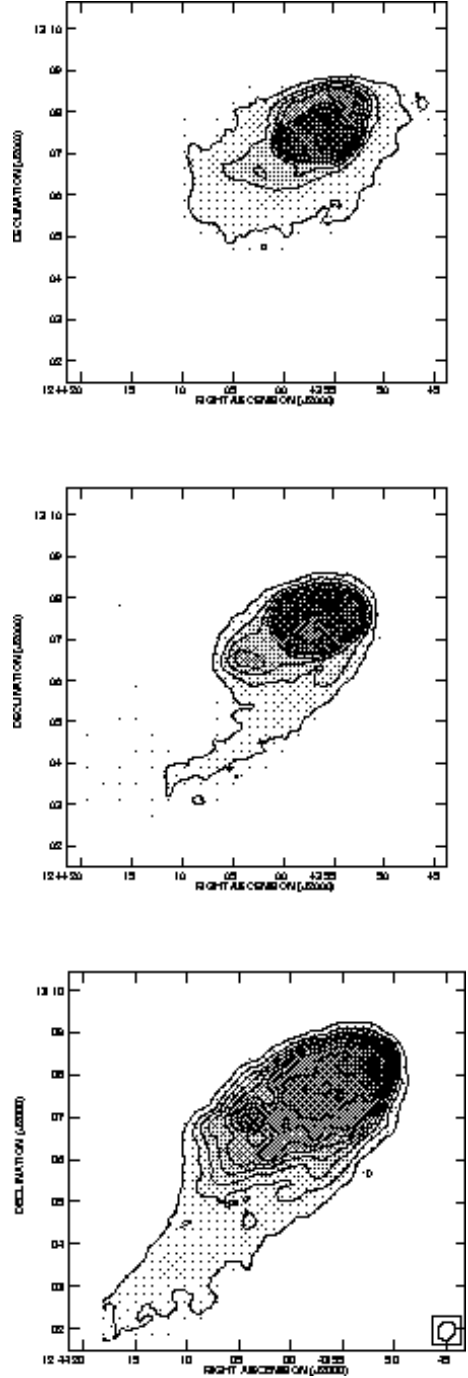}}
        \caption{Top panel: model distribution of the H{\sc i} gas
        for a model of a gravitational interaction alone.
        The contours correspond to 
        (1, 5 ,9 ,13, 17, 21, 25)$\times$10$^{19}$~cm$^{-2}$.
        Middle panel: distribution of the H{\sc i} gas
        of the model of a gravitational interaction together with 
        constant ram pressure.
        Bottom panel: H{\sc i} distribution of NGC~4654 (Phookun \& Mundy 1995)
        with the same contour levels.
        } \label{fig:grav_dist}
\end{figure}

Fig.~\ref{fig:grav_dist} shows the H{\sc i} distribution of
NGC~4654 (bottom panel) together with the model gas distributions convolved
to the beamsize of the 21--cm--line observations (25$''$).
The top panel shows the gas distribution of the model of a gravitational 
interaction alone, the middle panel shows that a gravitational
interaction together with constant ram pressure.
For both models the asymmetry of the gas distribution within the 
optical radius along the major axis can be reproduced.
The observed local maximum in the north--west as well as the extended region 
in the south--east are clearly visible in both model snapshots.
The simulations using only ram pressure as perturbation could
not reproduce these features (see Fig.~\ref{fig:HI_distributions}).

The very extended south--eastern tail can only be reproduced by the
model including ram pressure.

\begin{figure}
        \resizebox{8cm}{!}{\includegraphics{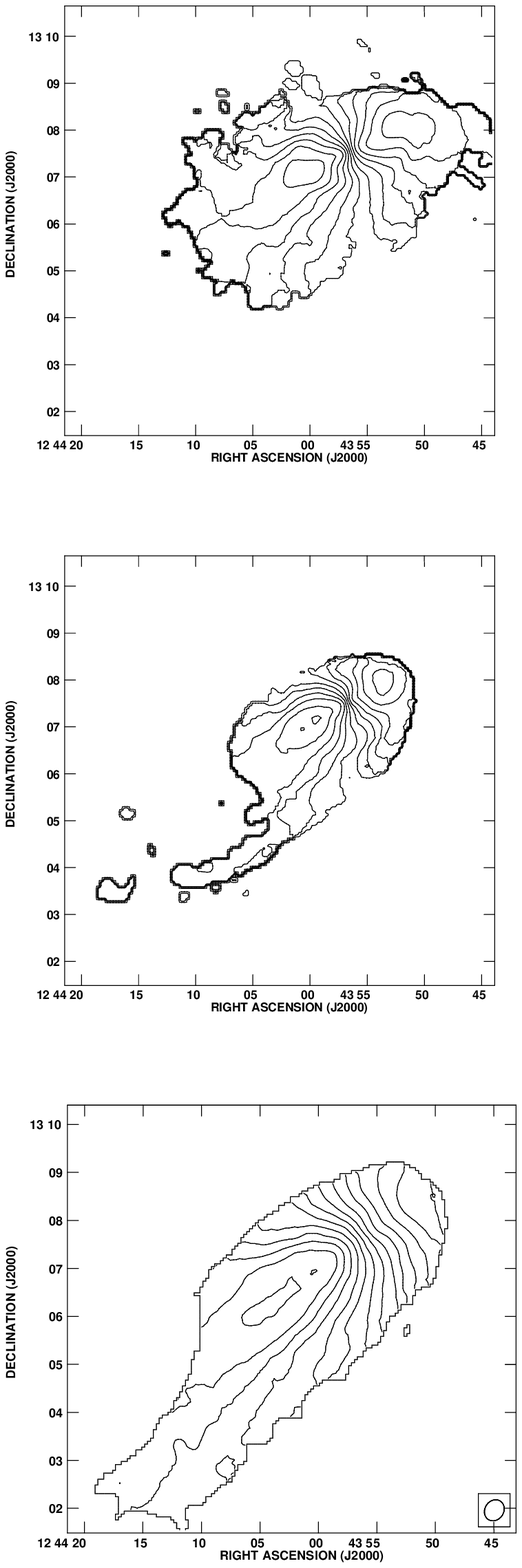}}
        \caption{Top panel: model H{\sc i} velocity field of a gravitational
          interaction alone. The contours, 
          in km\,s$^{-1}$, are from 900 (north--west)
          to 1160 (south--east) in steps of 20 for both velocity fields.
          Middle panel: gas velocity field of the model of a gravitational
          interaction together with constant ram pressure 
          with the same contour levels.
          Bottom panel : H{\sc i} velocity field of NGC~4654 
          (Phookun \& Mundy 1995) with the same contour levels.
        } \label{fig:grav_vel}
\end{figure}

The velocity fields of the two models and the observed velocity
field can be seen in Fig.~\ref{fig:grav_vel}. 

\begin{figure}
        \resizebox{\hsize}{!}{\includegraphics{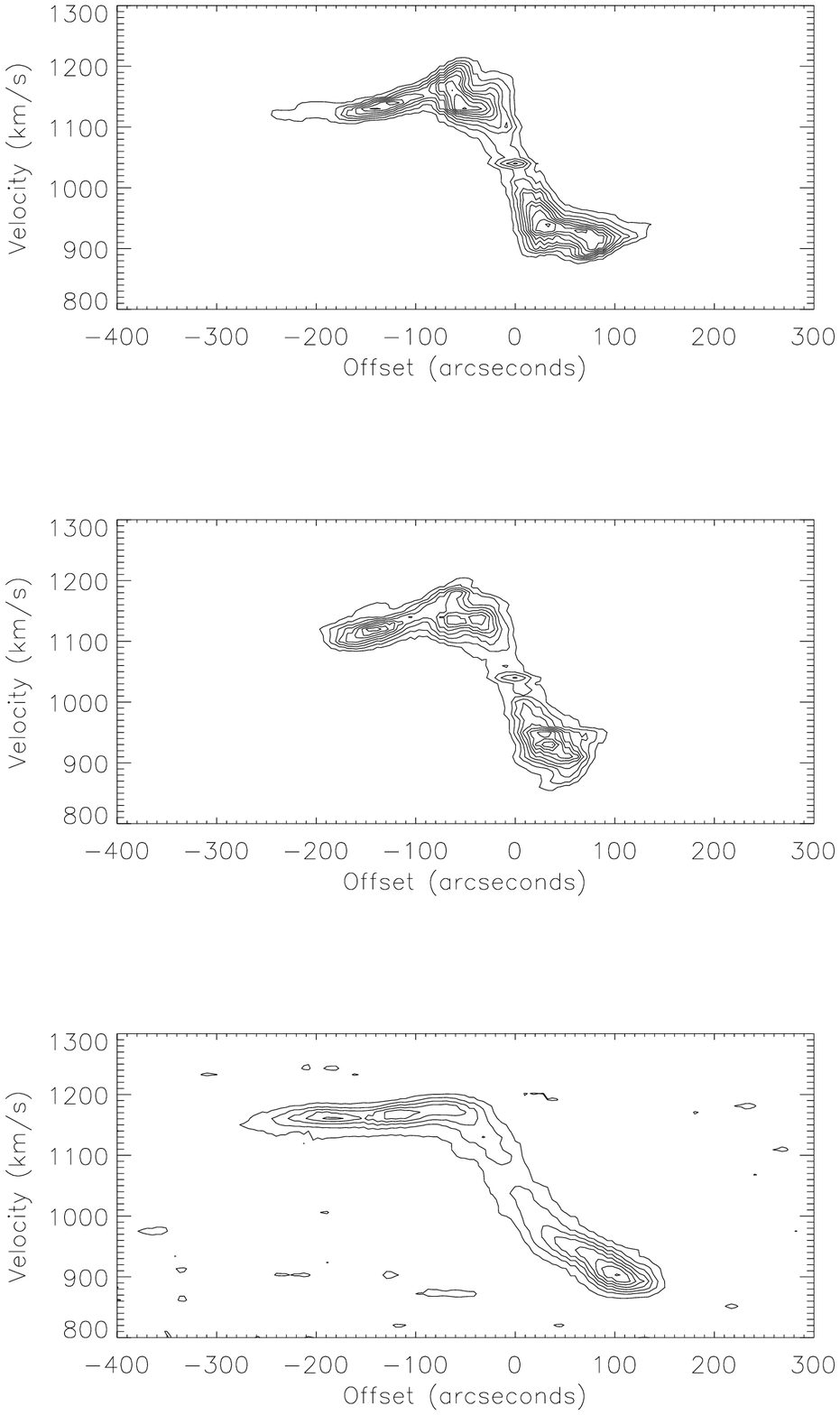}}
        \caption{Position--velocity cut through the H{\sc i} cube along the major axis.
        Top panel: model of a gravitational interaction alone.
        Middle panel: model of a gravitational interaction together
        with constant ram pressure.
        Bottom: H{\sc i} (Phookun \& Mundy 1995).
        } \label{fig:major}
\end{figure}

\begin{figure}
        \resizebox{\hsize}{!}{\includegraphics{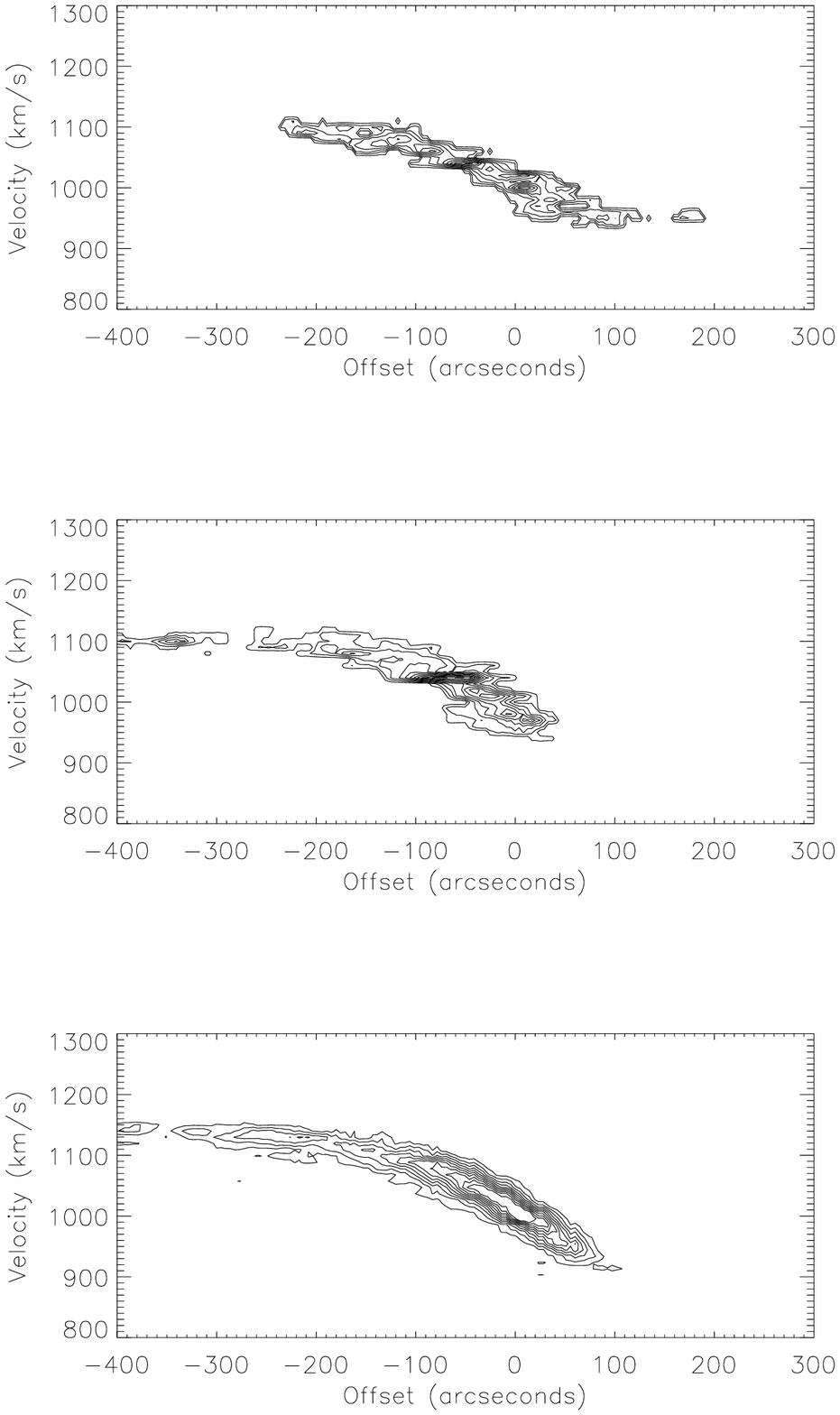}}
        \caption{Position--velocity cut through the H{\sc i} cube 
	along the extended tail.
        Top panel: model of a gravitational interaction alone.
        Middle panel: model of a gravitational interaction together
        with constant ram pressure.
        Bottom: H{\sc i} (Phookun \& Mundy 1995).
        } \label{fig:tail}
\end{figure}

\begin{figure}
        \resizebox{\hsize}{!}{\includegraphics{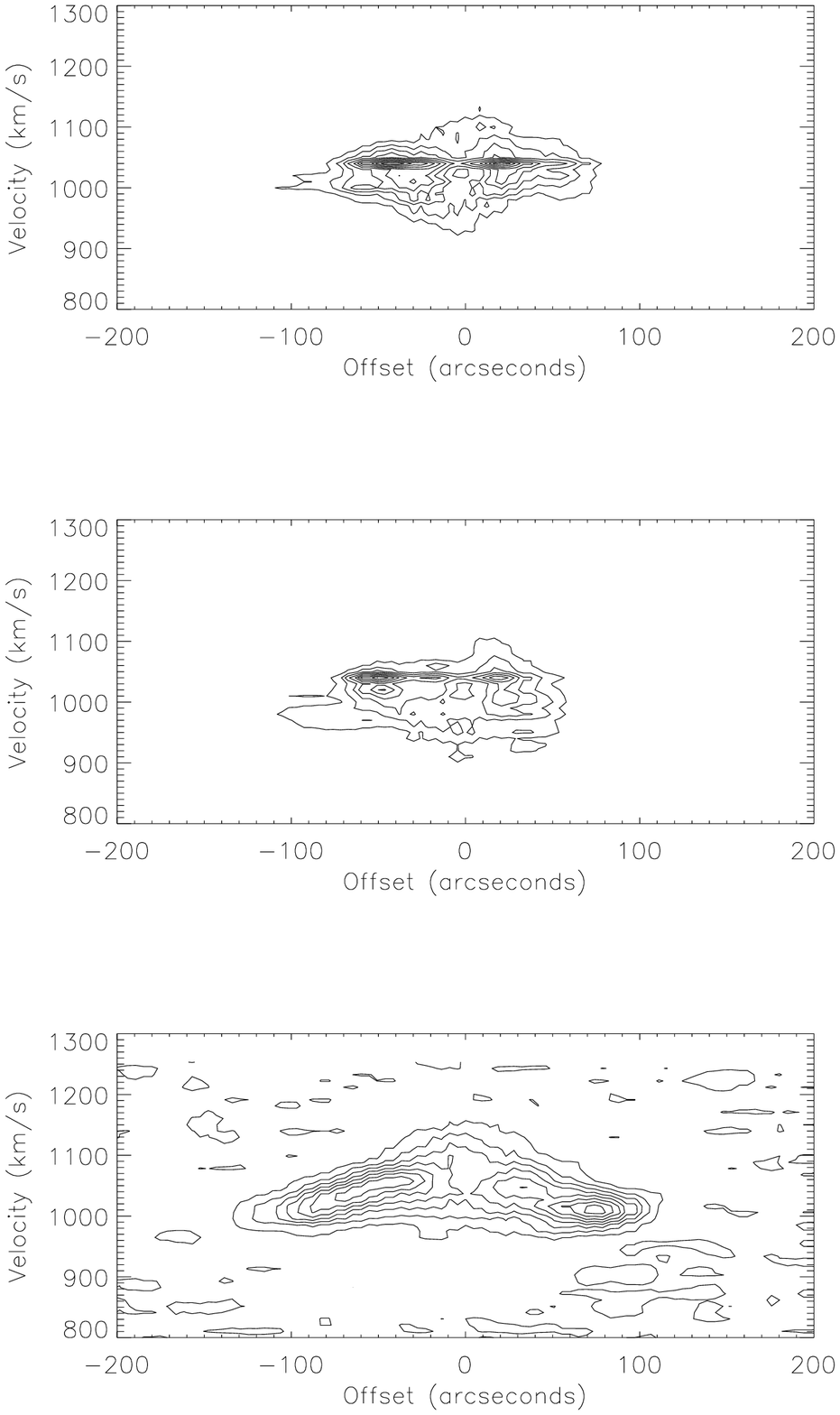}}
        \caption{Position--velocity cut through the H{\sc i} cube along the minor axis.
        Top panel: model of a gravitational interaction alone.
        Middle panel: model of a gravitational interaction together
        with constant ram pressure.
        Bottom: H{\sc i} (Phookun \& Mundy 1995).
        } \label{fig:minor}
\end{figure}

In order to compare the characteristic structure of the velocity fields, 
the position--velocity diagrams
along (i) the major axis, (ii) the tail, and (iii) the minor axis 
are shown in Fig.~\ref{fig:major}, \ref{fig:tail}, and
\ref{fig:minor}, respectively.
The south-eastern side (negative offsets) of the model cuts along the major axis 
are very similar for both models, i.e. the dynamics of this part of the disk are 
not affected by ram pressure. The model cuts show a maximum at an 
offset of $\sim$-50$"$. The rotation curve then decreases slightly. 
The observed rotation curve does not show a maximum. It reaches the plateau of 
constant rotation velocity at an offset of -40$"$.
The north--western side (positive offsets) of the galaxy shows a slightly rising 
rotation curve for the model including only a gravitational interaction. In contrast,
the model of the mixed interaction (gravitational and ram pressure) leads to a linear
rising rotation curve in the north-west.
Again, at the north--western edge the rotation curve is falling 
due to the velocity dispersion of the gas within the disk, which is higher than
that of the observed galaxy, because there is no explicit cooling mechanism
included in the model (cf. Sect.~\ref{sec:compare}).

Since the model of a mixed interaction and observations show a constant rotation curve 
at one side and a rising rotation curve at the other side, I conclude that 
the main characteristics of the observed rotation curve are reproduced by this model.

The p--V cuts along the tail reflect the structure of the gas distributions
(Fig.~\ref{fig:grav_dist}). The gas distribution of the model 
without ram pressure extends further to the north--west (positive offsets) 
than the one including ram pressure. At the opposite side the situation is 
reversed. The model including ram pressure reproduces nicely the observed
p--V cut along the tail.
The p--V cuts of both models along the minor axis are very similar.
They show a smaller but visible degree of asymmetry compared to
the observed asymmetry.

\section{Discussion}

The main differences between the two models using ram pressure as the only
perturbation are that (i) the disk is heated during the ICM--ISM interaction
and (ii) the stellar distribution reacts to the heavily perturbed gas distribution.
Whereas (ii) is a robust result, the disk heating might be artificial,
because of the discreteness of the model and because there is no gas cooling 
included in the model. Thus, the decrease of the gas column density in the
extended tail has to be interpreted with caution. If one wants to compare
this kind of models to observations of other galaxies, both models have to be
taken into account. The situation is still more complicated by the fact that
at the outer regions of the gas distribution evaporation effects might play
an important r\^{o}le that can in principle lower the gas column density.

The comparison between the models and observations shows important differences.
Whereas the model using ram pressure as the only perturbation can explain
the observed asymmetric rotation curve and the extended, low surface 
density tail of NGC~4654, it cannot reproduce the
observed velocity structure of the extended, south--eastern tail.
Nevertheless, it is worth noting that a past, strong ram pressure event 
can produce a tail--like structure even 800~Myr after the galaxy's closest
passage to the cluster center. Since we observe a reaction of the stellar
disk to the strongly perturbed gas distribution, the perturbation might be
longer--lived than expected if one takes only the timescale of gas diffusion 
into account. Moreover, it is not possible to form a bar and asymmetric
spirals as observed with ram pressure as the only perturbation.

The model using a gravitational interaction is able to reproduce the observed
stellar distribution of NGC~4654. The chosen orbit of the perturbing galaxy, 
NGC~4639, is not unique. 
It has been chosen such that (i) the position of the perturber
and its velocity are close to those observed for NGC~4639 and (ii) to
reproduce the asymmetry of the stellar distribution. The aim of this work is
not to explain all details of the observations, but to reproduce the main
characteristics of the observations. It becomes clear that only the
combination of a gravitational interaction and ram pressure can account
for all characteristics. I want to stress here that it is important to compare
the gas distribution and the gas kinematics. In the case of NGC~4645 it is the
gas kinematics that discriminate between the models.
Since the tidal interaction pulls gas to larger galactic radii,
only a small amount of ram pressure is needed to produce the
very extended gas tail.

Despite the fact that only the mixed perturbation (gravitational and ram
pressure) can explain all characteristics of the observations, there are
two potential problems with this model:
\begin{itemize}
\item
is an impact parameter between the two galaxies of $\sim$20~kpc realistic? 
In this case the disk of NGC~4639 has to be almost parallel to that of 
NGC~4654 during the encounter to avoid an ISM--ISM interaction, which would
complicate the scenario.
This  can be verified by making a model using two galaxy models
that include a collisional and a non--collisional component.
This introduces a further open parameter, i.e. the inclination angle
of disk of NGC~4639 with respect to the disk of NGC~4654.
\item
In the restframe of NGC~4654, NGC~4639 has a velocity in the plane of the sky
of $v_{\rm N4639} \sim 145$~km\,s$^{-1}$ towards the north--west.
With respect to the cluster both galaxies have a velocity of
$\sim$1000~km\,s$^{-1}$ mainly towards the cluster center (M87).
Thus the atomic gas outside the optical radius of NGC~4639 should
be affected by ram pressure in the same way as the gas of NGC~4654,
i.e. it should be mainly located to the south--east of the galaxy center.
Warmels (1988) observed indeed extended gas at the east of NGC~4639.
However, this is not observed by Cayatte et al. (1990). They found an 
external emission region in the south--west of the galaxy center.
Moreover Phookun \& Mundy (1995) did not detect any emission in the
east outside the optical radius of NGC~4639.
Since NGC~4639 is located at the edge of the primary beam of
Cayatte et al. (1990) and Phookun \& Mundy (1995) their results
have to be confirmed by observations especially dedicated to NGC~4639. 
\end{itemize}

\section{Summary and conclusions}

Numerical models including ram pressure and gravitational interaction 
are compared to high--sensitivity H{\sc i} observations of the spiral galaxy 
NGC~4654 in the Virgo cluster. This galaxy shows a very extended, low surface
density tail that does not show clear signs of rotation.

Four different models are presented, in order to investigate the origin
of this tail:
\begin{itemize}
\item
fixed gravitational potential of the galaxy, ram pressure is included,
\item
stellar content and dark halo simulated by a non--collisional component,
ram pressure is included,
\item
gravitational interaction, 
\item
gravitational interaction, ram pressure is included.
\end{itemize}

\subsection{Ram pressure as the only perturbation}

The temporal ram pressure profile
corresponds to a realistic galaxy orbit within the cluster. 
Among the simulations of Vollmer et al. (2001) with different
inclination angles $i$ between the disk and the orbital plane,
one simulation was chosen, which reproduces the observed H{\sc i}
deficiency and the H{\sc i} distribution. Limits for these parameters are: 
5000~cm$^{-3}$(km\,s$^{-1}$)$^{2} < p_{\rm ram} 
< 10000$~cm$^{-3}$(km\,s$^{-1}$)$^{2}$ and $0^{\rm o} < i < 10^{\rm o}$. 
The timestep of the snapshot compared to the observations
is roughly given by the projected distance of NGC~4654 to the cluster center
divided by 1700~km\,s$^{-1}$, which represents a typical velocity
(with respect to the cluster mean velocity) of a
mildly stripped galaxy in the core of the Virgo cluster (Vollmer et al. 2001).
With the given PA and inclination angle, the azimuthal angle 
has to be chosen such that the radial velocity of the
model galaxy equals approximately that of NGC~4654.
These are very severe constraints for the model, reducing the number of
possible snapshots to a few.

\subsection{Gravitational interaction}

The parameters for the gravitational interaction were chosen such
that (i) the observed position and radial velocity of NGC~4639 are
reproduced and (ii) the observed asymmetry of the stellar content of
NGC~4654 is reproduced. Asymmetric spiral arms without disrupting the galaxy
can only be obtained with a very close, retrograde encounter.

\subsection{Results}

The following results were obtained from the different models:
\begin{enumerate}
\item
The model that uses ram pressure as the only perturbation shows
a tail up to 800~Myr after a strong ram pressure event, i.e. after
the galaxy's passage through the cluster center.
\item
During the ICM--ISM interaction the non--collisional component of the 
galaxy is heated resulting in a tail of low surface density.
In the case of a fixed gravitational potential the tail has a
much higher surface density, because of the lower scale height
of the disk's gravitational potential.
\item
An edge--on ICM--ISM interaction can produce an asymmetric stellar
distribution and thus an asymmetric rotation curve.
\item
An edge--on ICM--ISM interaction cannot produce bar.
\item
The model using a gravitational interaction as the only perturbation
can reproduce the observed asymmetry of the stellar content of NGC~4654.
\item
The model using a gravitational interaction as the only perturbation
cannot reproduce the observed, extended gas tail of NGC~4654.
\item
Only the model of a mixed perturbation (gravitational and ram pressure)
can account for all observed properties (gas and stars) of NGC~4654.
In the case of a past tidal interaction only a small amount of ram pressure 
is needed to form the observed asymmetries of the gas distribution and
velocity field.
\item
Only the comparison with the gas distribution {\it and} velocity field
can discriminate between the models.
\end{enumerate}

I thus conclude that NGC~4654 has suffered most probably a
tidal interaction with its companion NGC~4639 $\sim$500~Myr ago.
It is now entering the cluster with a velocity of $\sim$1000~km\,s$^{-1}$
mainly to the west. It is now experiencing a small ram pressure of the
order $p_{\rm ram} \sim 200$~cm$^{-3}$km\,s$^{-1}$ that is responsible
for the observed extended, low surface density gas tail.

\begin{acknowledgements}
The author would like to thank L. Hernquist for making his program that
generates initial conditions available, C. Balkowski for fruitful discussions,
E.M. Berkhuijsen for reading the article, and
the NCSA Astronomy Digital Image Library (ADIL) for providing the data cube 
for this article. A special thanks to the referee, J. van Gorkom, who
helped me to improve this article substantially.
\end{acknowledgements}


\begin{thebibliography}{}

\bibitem{q1} Allen C., Santill\'an A. 1991, RMAA, 22, 255
\bibitem{q2} Byrd G., Freeman T., Howard S. 1993, AJ, 105, 477
\bibitem{q3} Boselli A., Tuffs R.J., Gavazzi G., Hippelein H., \& Pierini D. 1997, A\&AS, 121, 507
\bibitem{q4} Cayatte V., van Gorkom J.H., Balkowski C., Kotanyi C. 1990, AJ, 100, 604
\bibitem{q5} Cayatte V., Kotanyi C., Balkowski C., van Gorkom J.H. 1994, AJ, 107, 1003 
\bibitem{q6} de Vaucouleurs G., de Vaucouleurs A., Corwin H.G., Buta R.J., Paturel G., Fouqu\'e P. 1991, Third Reference Catalogue of Bright Galaxies,(New York:Springer)(RC3)
\bibitem{q7} Guharthakurta P., van Gorkom J.H., Kotanyi C.G., Balkowski C. 1988, AJ, 96, 851
\bibitem{q8} Gunn J.E., Gott J.R. 1972, ApJ, 176, 1
\bibitem{q9} Hernquist L. 1993, ApJS, 86, 389
\bibitem{q9a} Huchtmeier W.K. \& Richter O.-G. 1989, A General Catalogue of H{\sc i} observations of Galaxies (New York: Springer-Verlag)
\bibitem{q10} Kenney J.D., Young J.S. 1989, ApJ, 344, 171 
\bibitem{q11} Phookun B., Mundy L.G. 1995, ApJ, 453, 154
\bibitem{q12} Sperandio M., Chincarini G., Rampazzo R., de Souza R. 1995,
  A\&AS, 110, 279 
\bibitem{a13} Springel V., Yoshida N., \& White D.M. 2001, NA, 6, 79
\bibitem{q14} Thomasson M., Donner K.J., Sundelius B., Byrd G.G., Huang T.-Y., Valtonen M.J. 1989, A\&A, 211, 25
\bibitem{q15} Vollmer B., Cayatte V., Balkowski C., Duschl W.J. 2001, ApJ, 561, 708
\bibitem{q16} Warmels R.H. 1988 A\&AS, 72, 57 
\bibitem{q17} Wiegel W. 1994, Diploma Thesis, University of Heidelberg

\end{thebibliography}
\end{document}